\documentclass[a4paper]{article}

\usepackage{INTERSPEECH2021}

\usepackage{bm,amsmath,amssymb,mathrsfs,amsfonts}
\usepackage{caption}
\usepackage{scalefnt}
\usepackage{multirow}
\usepackage{here}
\usepackage{booktabs}
\usepackage{enumitem}
\usepackage{type1cm}
\usepackage{cite}
\usepackage{url}
\usepackage{algorithm}
\usepackage{algpseudocode}
\usepackage{arydshln}
\usepackage{color}

\newcommand{\argmax}{\mathop{\rm argmax}\limits}
\newcommand{\maxcustom}{\mathop{\rm max}\limits}

\usepackage{pifont}%
\newcommand{\cmark}{\ding{51}}%
\newcommand{\xmark}{\ding{55}}%

\newcommand{\xblock}{x^{m}}
\newcommand{\hiddenblock}{h^{m}}
\newcommand{\hiddenblocklast}{h^{m}_{|h^{m}|-1}}
\newcommand{\nbeam}{B}
\newcommand{\nframe}{t}
\newcommand{\largenframe}{N_{\rm sg}}
\newcommand{\tblock}{T_{\rm block}}
\newcommand{\tblocksub}{T'_{\rm block}}
\newcommand{\omegaplus}{\Omega_{+}}
\newcommand{\omegaminus}{\Omega_{-}}
\newcommand{\omeganext}{\Omega_{\rm next}}
\newcommand{\omegaend}{\Omega_{\rm eos}}
\newcommand{\omegasession}{\Omega_{\rm session}}
\newcommand{\nblank}{n_{\varnothing}}
\newcommand{\largenblank}{N_{\varnothing}}
\newcommand{\ratiolen}{R_{\rm len}}
\newcommand{\pspike}{P_{\rm spike}}

\newcommand{\lambdase}{\lambda_{\rm se}}
\newcommand{\lambdaqua}{\lambda_{\rm qua}}
\newcommand{\lambdasync}{\lambda_{\rm sync}}

\title{VAD-free Streaming Hybrid CTC/Attention ASR for Unsegmented Recording}
\name{Hirofumi Inaguma, Tatsuya Kawahara}
\address{
  Graduate School of Informatics, Kyoto University, Kyoto, Japan}
\email{\{inaguma,kawahara\}@sap.ist.i.kyoto-u.ac.jp}

\begin{document}

\maketitle
\begin{abstract}
In this work, we propose novel decoding algorithms to enable streaming automatic speech recognition (ASR) on unsegmented long-form recordings without voice activity detection (VAD), based on monotonic chunkwise attention (MoChA) with an auxiliary connectionist temporal classification (CTC) objective.
We propose a \textit{block-synchronous} beam search decoding to take advantage of efficient batched output-synchronous and low-latency input-synchronous searches.
We also propose a VAD-free inference algorithm that leverages CTC probabilities to determine a suitable timing to reset the model states to tackle the vulnerability to long-form data.
Experimental evaluations demonstrate that the block-synchronous decoding achieves comparable accuracy to the label-synchronous one.
Moreover, the VAD-free inference can recognize long-form speech robustly for up to a few hours.
\end{abstract}

\section{Introduction}
Recent progress of end-to-end (E2E) automatic speech recognition (ASR) enables us to build competitive systems to conventional hybrid systems with much smaller development efforts.
For live streaming applications, frame-synchronous models such as connectionist temporal classification (CTC)~\cite{ctc_graves} and RNN transducer (RNN-T)~\cite{rnn_transducer} are promising approaches because of the robustness for long-form speech~\cite{narayanan2019recognizing,chiu2019comparison}.
Attention-based encoder-decoder (AED)~\cite{chorowski2015attention,chan2016listen} have shown outstanding performances in the offline task~\cite{s2s_comparison_google,s2s_comparison_baidu,li2020comparison} and have been intensively investigated for streaming extensions~\cite{hard_monotonic_attention,mocha,cif,tsunoo2020streaming}.
Among them, monotonic chunkwise attention (MoChA)~\cite{mocha} is attractive because of the monotonic constraint of alignments and linear-time decoding complexity at test time.
The notable advantages of MoChA over frame-synchronous models are faster decoding thanks to label-wise predictions~\cite{inaguma2021alignment} and availability of large vocabularies because of lower memory consumption.
Moreover, MoChA does not have to expand hypotheses over a silence region because it treats silence in the internal attention module.

However, the generalization capability of the AED models to long-form speech is poor~\cite{chiu2019comparison,dong2020comparison}, and how to mitigate this problem is still an open question.
Several methods have tackled this problem by incorporating alignment information to the training as supervision~\cite{inaguma2020streaming,inaguma2020_ctcsync,inaguma2021alignment}, window-based overlapped offline inference~\cite{chiu2019comparison,kang2021partially}, modifying LSTM encoder states~\cite{narayanan2019recognizing}, and adopting new architecture~\cite{cif,dong2020comparison}.
It is also a common practice to segment long-form audio with a separate voice activity detection (VAD) model in advance~\cite{hughes2013recurrent}. CTC can also be used for that purpose~\cite{yoshimura2020end,fujita2020end,li2021long}.
Joint end-pointing was also investigated for RNN-T in the voice search task~\cite{chang2019joint,li2020towards,mahadeokar2020alignment}.

In this work, we propose novel decoding algorithms to recognize speech of unlimited length in a streaming way with MoChA trained jointly with an auxiliary CTC loss~\cite{kim2017joint,hybrid_ctc_attention}.
Instead of pursuing better generalization to long-form speech from a training perspective, we seek a solution to find a suitable timing to reset the model states from a decoding perspective.
Firstly, we propose a \textit{block-synchronous} beam search decoding, in which the advantages of breadth- and best-first searches are taken to achieve efficient batched inference and low display latency.\footnote{We define display latency as a delay to emit tokens caused by the search procedure and distinguish it from the emission latency~\cite{sak2015acoustic_asru,inaguma2020streaming,wang2020low,mahadeokar2020alignment,yu2021fastemit,yu2021dualmode,inaguma2021alignment}, which is caused by the E2E model training. We have already tackled to reduce the emission latency of MoChA in~\cite{inaguma2021alignment,inaguma2021stableemit}.}
We allow continuing search within a block by relaxing the label-synchronous hypothesis pruning to consider (potentially) various lengths of candidates in the beam given a partial observation.
Secondly, we propose a VAD-free streaming inference algorithm leveraging CTC probabilities to determine a timing to reset the states.
Instead of performing audio segmentation before ASR, our method recognizes all speech, including silence frames, and therefore the ASR model does not have to wait for the segmentation to be completed.
Moreover, the unified framework is suitable for context management and on-device applications.
Although our base model is MoChA, the proposed VAD-free inference can also be applied to any streaming ASR model trained jointly with the CTC objective.

Experimental evaluations on English and Japanese lecture corpora demonstrate that the block-synchronous decoding achieves comparable accuracy to the label-synchronous decoding and even outperforms it in some cases.
We also show that the VAD-free inference does not degrade accuracy so much without the ground-truth segmentation and achieves better performance than cascading an external VAD model.

\vspace{-1mm}
\section{Streaming Hybrid CTC/Attention ASR}\label{sec:background}

MoChA extended hard monotonic attention (HMA)~\cite{hard_monotonic_attention} by equipping an additional chunkwise soft attention module restricted to local $w$ frames.
To generate the $i$-th token with a linear-time complexity at test time, HMA introduces a discrete decision $z_{i,j}\in\{0,1\}$ ($j$: encoder time index) and samples it from a Bernoulli random variable, $\mbox{Bernoulli}(p_{i,j})$, where $p_{i,j} \in [0,1]$ is a selection probability as a function of encoder and decoder outputs.
To enable the backpropagation training, the expected alignment score $\alpha_{i,j}$ is calculated with $p_{i,j}$ by considering all alignment paths as 
\begin{eqnarray}
\alpha_{i,j} &=& p_{i,j}\bigg((1-p_{i,j-1})\frac{\alpha_{i,j-1}}{p_{i,j-1}}+\alpha_{i-1,j}\bigg). \label{eq:hma_alpha}
\end{eqnarray}
The chunkwise attention score for a context vector is calculated with $\alpha_{i,j}$ at training time and with discrete indices at test time.
We also apply \textit{StableEmit}~\cite{inaguma2021stableemit} to reduce the emission latency by multiplying $p_{i,j}$ in Eq.~\eqref{eq:hma_alpha} by a constant factor $1 - \lambdase$ ($\lambdase > 0$) during training.

Applying an auxiliary CTC loss $\mathcal{L}_{\rm ctc}$ on top of the encoder of AED models is effective in encouraging the decoder to learn a monotonic alignment~\cite{kim2017joint,hybrid_ctc_attention,online_hybrid_ctc_attention,online_hybrid_ctc_attention_taslp2020}.
We also introduce a quantity loss $\mathcal{L}_{\rm qua}$~\cite{inaguma2020streaming} or a CTC-synchronous training (CTC-ST) loss $\mathcal{L}_{\rm sync}$~\cite{inaguma2020_ctcsync} to improve the performance and reduce the emission latency~\cite{inaguma2021alignment,inaguma2021stableemit}.
The total objective $\mathcal{L}_{\rm total}$ is formulated as
\begin{multline}
\mathcal{L}_{\rm total} = (1- \lambda_{\rm ctc}) \mathcal{L}_{\rm mocha} + \lambda_{\rm ctc} \mathcal{L}_{\rm ctc} \\ 
+ \lambdaqua \mathcal{L}_{\rm qua} + \lambdasync \mathcal{L}_{\rm sync}, \label{eq:total_loss}
\end{multline}
where $\lambda_{*}$ is a corresponding task weight.
Unlike~\cite{hybrid_ctc_attention}, we do not perform joint CTC decoding during beam search because they were not helpful in our experiments.

\vspace{-1mm}
\section{Efficient block-synchronous decoding}\label{sec:block_sync}
When using MoChA, beam search decoding is conducted in a label-synchronous way (i.e., breadth-first search) at test time to find the most probable output sequence.
However, active hypotheses in the current beam are pruned after the expansion when and only when (1) all the hypotheses find the next token boundaries (i.e., $j$ s.t. $z_{i,j}=1$) or (2) their pointers to the encoder outputs reach the last encoder output observed so far.
In other words, all the active hypotheses must have the same output sequence length at each output step.
Therefore, the search cannot proceed forward if some active hypotheses fail to detect the next boundaries correctly, even when a new acoustic observation comes in.
Moreover, when applying subword tokenization (e.g., byte pair encoding (BPE)~\cite{sennrich2015neural}) to word sequences, hypotheses in the beam could have different lengths even when they correspond to the same word sequence.
This problem becomes more serious when recognizing long-form speech, resulting in a non-negligible recognition delay in the online streaming scenario.
So our goal is to continue sequence generation as long as some of the active hypotheses detect the subsequent token boundaries over the current acoustic observation.

\vspace{-1mm}
\subsection{Proposed algorithm}
To perform beam search with a minimal display latency given a partial acoustic observation, we propose an efficient \textit{block-synchronous} beam search decoding for MoChA, which relaxes the constraint of the label-synchronous hypothesis pruning.
The block-synchronous decoding combines the advantages of breadth- and best-first searches, efficient batched computation~\cite{seki2019vectorized} and small display latency.
The proposed algorithm is shown in Algorithm~\ref{algo:blocksync}.
Given the $m$-th input block $\xblock$ of a fixed length $\tblock$ [10ms], we perform the breadth-first search over the corresponding encoder outputs $\hiddenblock$ of length $\tblocksub$ ($< \tblock$).
Unlike label-synchronous decoding, however, active hypotheses in the current beam are forcibly pruned no matter whether all active hypotheses detect the next boundaries in $\hiddenblock$.
The search in the $m$-th block continues until (1) none of the active hypotheses find any further boundary in $\hiddenblock$ (\textit{line:\ref{algo:no_boundary_all}}) or (2) the number of generated tokens in $\hiddenblock$ surpasses $U_{\rm max}(=\tblocksub \times \ratiolen)$.\footnote{$\ratiolen$ is a hyperparameter to control the maximum output length in each block, but we did not observe token repetition.}

Let $\omegaplus$ be a set of hypotheses having a possibility to detect the next boundary in $\hiddenblock$.
Hypotheses in $\omegaplus$ are added to $\omegaminus$ without prefix expansion if any next boundary is not detected in $\hiddenblock$ (\textit{line:\ref{algo:no_boundary_each}}).
In this case, we allow to generate $\langle$eos$\rangle$ only because of MoChA's behavior (see Algorithm 1 in~\cite{mocha}).
Otherwise, it is expanded by the top-$k$ tokens and is added to $\omeganext$ (\textit{line:\ref{algo:blocksync_expand}}).
When $\langle$eos$\rangle$ is generated, the hypothesis is added to a complete hypothesis set $\omegaend$ instead (\textit{line:\ref{algo:blocksync_eos}}).
Pruning is conducted over $\omeganext$\footnote{This is more effective than pruning over $\omeganext \cup \omegaminus$.} with a beam width $\nbeam$ at every output step (\textit{line:\ref{algo:blocksync_prune}}).
To avoid biasing to shorter hypotheses because of the monotonic decrease of sequence-level log probabilities, we normalize them by the current output sequence length (\textit{length normalization}~\cite{murray-chiang-2018-correcting}).

The entire search process for an utterance or a session is finalized when all pointers to $\hiddenblock$ in $\omegaplus$ reach the last encoder output,  $\hiddenblocklast$.
The details will be described in Section~\ref{sec:vad_free}.
The search is equivalent to frame-synchronous and label-synchronous decoding by setting $\tblocksub$ to 1 and $\infty$, respectively.
Therefore, the proposed algorithm is a generalized form of both search methods. 

Concurrently to this work, streaming block-synchronous decoding with an offline Transformer decoder was proposed in~\cite{tsunoo2020streaming}.
They determined to move to the next block when generating $\langle$eos$\rangle$ at the current block and introduced complicated heuristics to avoid token repetition.
The decoding complexity was quadratic of the input length because of incremental decoding.
In contrast, our method can move to the next block without generating $\langle$eos$\rangle$ in each block because MoChA has a function of detecting a token boundary at \textit{each frame}.
This also saves a computation of the softmax normalization in the output layer.
Moreover, the decoding complexity of our method is linear, and we do not perform joint CTC decoding.

\begin{algorithm}[t]
\caption{Block-synchronous beam search decoding with MoChA at the $m$-th block}
\footnotesize
\begin{algorithmic}[1]
\Function{BlockSync}{$\hiddenblock, \omegaplus, \omegaend, \nbeam, \ratiolen$}
    \State $U_{\rm max} \gets |\hiddenblock| \times \ratiolen$, $\omegaminus \gets \{\}$

    \For {$i = 1, \ldots, U_{\rm max}$}
        \If {$|\omegaplus| = 0$}
                \State \textbf{break} {\color{blue} \Comment{Move to the next block}} \label{algo:no_boundary_all}
            \EndIf \label{algo:no_boundary_end}

        \State $p_{{\rm mocha}, i} = {\rm Decoder}(\omegaplus, [h^{m-1}_{-(w-1):}; \hiddenblock])$  \label{algo:decoder_update}
        \State $score = \log p_{{\rm mocha}, i} + \lambda_{\rm lm} \log p_{{\rm lm}, i}$ {\color{blue} \Comment{Normalize by length}}

        \State $\omeganext \gets \{\}$
        \For {$y$ in $\omegaplus$}
            \If {$\sum_{j} z_{i,j} = 0$}
                \State add $y$ to $\omegaminus$ {\color{blue} \Comment{No boundary detected}} \label{algo:no_boundary_each}
            \EndIf

            \For {$k \in \mathcal{V}$}
                    \If {$k = \langle$eos$\rangle$}
                        \State add $y$ to $\omegaend$ \label{algo:blocksync_eos}
                    \Else
                        \State add $y + [k]$ to $\omeganext$ \label{algo:blocksync_expand}
                    \EndIf
             \EndFor
        \EndFor
        \State $\omegaplus \gets$ top-$\nbeam$ in $\omeganext$ \label{algo:blocksync_prune} {\color{blue} \Comment{Pruning}}
    \EndFor
    \State $\omegaplus \gets \omegaplus \cup \omegaminus$

    \State \Return ($\omegaplus, \omegaend$)
\EndFunction
\end{algorithmic}\label{algo:blocksync}
\end{algorithm}
\setlength{\textfloatsep}{2.0mm}  %

\vspace{-1mm}
\section{VAD-free streaming inference}\label{sec:vad_free}
\vspace{-1mm}
In the AED models, it is crucial to keep decoding contexts to improve the recognition performance because of the conditional dependency of output symbols.
However, it is required to exclude long samples from the training data to fit the GPU/TPU memory and perform efficient training.
Therefore, when recognizing long-form speech during inference, the models must generalize to unseen samples, but it is challenging in general~\cite{chiu2019comparison}.

\vspace{-1mm}
\subsection{Proposed algorithm}
To balance the limitation of the generalization capability to long-form speech and the effective context management, we determine a suitable timing to reset model states based on CTC probabilities.
We regard consecutive blank tokens ($\varnothing$) generated from the CTC branch as a silence region and use them to find a \textit{reset point}.
Unlike a CTC-based pre-segmentation in~\cite{yoshimura2020end}, our method does not perform the segmentation explicitly, i.e., we do not detect the onset.
Instead, we recognize all frames including long silence.
Therefore, we do not have to wait for the pre-segmentation to be completed to start recognition, leading to latency reduction.

The proposed algorithm is shown in Algorithm~\ref{algo:vad_free}.
We adopt the block-synchronous decoding in Section~\ref{sec:block_sync}.
We count the number of consecutive blank tokens $\nblank$ from the previous reset point and detect the next reset point when $\nblank$ surpasses a threshold $\largenblank$ (\textbf{condition 1}, \textit{line:\ref{algo:vad_free_cond1}}).
We also regard a weak non-blank spike whose probability is less than $\pspike$ as a blank token.
Note that the recognition continues until the end of the current block regardless of the result of the reset point detection.
Moreover, we also allow resetting the states when $\langle$eos$\rangle$ is generated (\textbf{condition 2}, \textit{line:\ref{algo:vad_free_cond2}}).
This is important for determining the reset point when enough silence is not found for a while.
Once a reset point is detected, we push the most probable hypothesis in $\omegaplus$ to a session-level hypothesis set $\omegasession$ and reset both decoder states and $\omegaplus$ (\textit{line:\ref{algo:vad_free_reset_beam_start}-\ref{algo:vad_free_reset_beam_end}}).
When using LSTM encoders, we also reset the encoder states.
To deal with speech frames around the block boundaries, we re-encode acoustic features in the previous block after the state reset and use the last states as the initial states in the current block (\textit{back-off initialization}).
When using Conformer encoders~\cite{gulati2020}, however, we do not have to reset the encoder states because they are agnostic to input offsets thanks to relative positional encoding and time-restricted self-attention.
To avoid frequent state resets, we introduce a safeguard in which the reset point detection is prohibited until the total number of input frames $\nframe$ from the previous reset point accumulates up to a threshold $\largenframe$ [10ms] (\textit{line:\ref{algo:vad_free_safeguard}}).
Moreover, LSTM LM states are carried over to the next block to provide useful contexts before the reset point.\footnote{We trained LSTM LM by carrying over the last state in a mini-batch to the initial state in the next mini-batch during LM training.}

\begin{algorithm}[t]
\caption{VAD-free streaming inference}
\footnotesize
\begin{algorithmic}[1]
\Function{Decode}{$x, \largenframe, \largenblank, \pspike, \ratiolen$}
    \State $\nframe \gets 0$, $\nblank \gets 0$
    \State $IsReset \gets False$
    \State $\omegasession \gets$ \{\}, $\omegaplus \gets \{\}$, $\omegaend \gets \{\}$

    \For {$m=1,\cdots, M$}
        \State $\hiddenblock \gets {\rm Encode}(\xblock, IsReset)$
        \State $\omegaplus, \omegaend \gets {\rm BlockSync}(\hiddenblock, \omegaplus, \omegaend, \nbeam, \ratiolen$)
        \State $\nframe \gets \nframe + |\xblock|$
        \If {$\nframe \geq \largenframe$} {\color{blue} \Comment{Safeguard}} \label{algo:vad_free_safeguard}
            
            \State $p^{\rm ctc} \gets {\rm CTC}(\hiddenblock)$
            \For {$j=1,\cdots, |\hiddenblock|$}
                \If {$\argmax_{k}p^{\rm ctc}_{j,k} = \varnothing$ or $\maxcustom_{k}p^{\rm ctc}_{j,k} < \pspike$}
                    \State $\nblank \gets \nblank + 1$ \label{algo:vad_free_blank_count}
                \Else
                    \State $\nblank \gets 0$
                \EndIf
                \If {$\nblank \geq \largenblank$}
                    \State {\color{red} $IsReset \gets True$}  {\color{blue} \Comment{Condition: 1}} \label{algo:vad_free_cond1}
                \EndIf
            \EndFor
            
            \If {$LastToken(\argmax(\omegaplus \cup \omegaend)) = \langle$eos$\rangle$}
                \State {\color{red} $IsReset \gets True$} {\color{blue} \Comment{Condition: 2}} \label{algo:vad_free_cond2}
            \EndIf
        \EndIf
        \If {$IsReset$} 
            \State $\omegasession.push(\argmax(\omegaplus \cup \omegaend))$ \label{algo:vad_free_reset_beam_start}
            \State $\nframe \gets 0$, $\nblank \gets 0$, $\omegaplus \gets$ \{\} {\color{blue} \Comment{Reset states}} \label{algo:vad_free_reset_beam_end}
            \State $IsReset \gets False$
        \EndIf
    \EndFor
    \State \Return $\omegasession$
\EndFunction
\end{algorithmic}\label{algo:vad_free}
\end{algorithm}
\setlength{\textfloatsep}{2.0mm}  %

\vspace{-1mm}
\section{Experimental evaluations}\label{sec:experiment}
\subsection{Experimental setup}\label{ssec:setup}
We used the TEDLIUM release v2 (TEDLIUM2)~\cite{tedlium} and the Corpus of Spontaneous Japanese (CSJ)~\cite{csj}.
TEDLIUM2 consists of about 210-hour English lecture speech.
CSJ consists of about 600-hour Japanese spontaneous academic lecture speech.
We combined three official test sets in CSJ: \textit{eval1}, \textit{eval2}, and \textit{eval3} to construct the \textit{test} set.
We extracted 80-channel log-mel filterbank coefficients computed with a 25-ms window shifted every 10ms with Kaldi~\cite{kaldi}.

We investigated three kinds of encoder architectures; unidirectional LSTM (UniLSTM), latency-controlled bidirectional LSTM (LC-BLSTM)~\cite{latency_controlled_blstm}, and latency-controlled Conformer (LC-Conformer) encoders~\cite{chen2020developing}.
The UniLSTM consisted of five layers of LSTM with 1024 units.
The LC-BLSTM had 512 units in both directions.
We set both the current and right block sizes to 40, i.e., 400ms.
The LC-Conformer encoder had the same architecture as Conformer (M)~\cite{gulati2020} with a kernel size of 15, while the number of layers was reduced from 16 to 12.
We used hierarchical downsampling~\cite{dong2019self} with the max-pooling having a stride of 2 at the last frontend CNN layer, 4th, and 8th Conformer blocks.
We also replaced batch normalization in each convolution module with layer normalization~\cite{inaguma2021stableemit}.
We adopted the masking strategy in~\cite{chen2020developing}, where lookahead frames were truncated in the same block, including the frontend CNN layers.
We set the left and current block sizes to 960ms and 320ms, respectively.
Therefore, the average algorithmic latency (AAL) of the UniLSTM, LC-BLSTM, and LC-Conformer encoders was 60ms, 660ms, and 160ms, respectively.
The decoder consisted of a single layer of LSTM with 1024 units.
We set a chunk size $w$ of MoChA to 4.

We applied CTC-ST~\cite{inaguma2020_ctcsync} to all LSTM models with $\lambdasync = 1.0$ in Eq.~\eqref{eq:total_loss}.
Moreover, StableEmit~\cite{inaguma2021stableemit} was applied to the UniLSTM and Conformer models on TEDLIUM2 with $\lambdase = 0.1$ and $\lambdaqua = 2.0$.
During inference, we used $\nbeam = 10$ with a four-layer LSTM LM.
We set ($\largenframe$, $\largenblank$, $\pspike$, $\ratiolen$) to (1600, 40, 0.1, 1.0).
The codes are publicly available.\footnote{\url{https://github.com/hirofumi0810/neural_sp}.}

\vspace{-1mm}
\subsection{Results}\label{ssec:result}
\vspace{-1mm}
\subsubsection{Utterance-level evaluation}
\vspace{-1mm}
We first compare the type of beam search decoding with the ground-truth segmentation in Table~\ref{tab:result_beam_search}.
Note that we did not use CTC probabilities to reset the model states here.
Using the UniLSTM encoder, we confirmed comparable WER with the block-synchronous decoding compared to the label-synchronous one on the TEDLIUIM2 \textit{dev} set while it was slightly degraded on the \textit{test} set.
Minimum WER training~\cite{prabhavalkar2018minimum} could mitigate the degradation.
In contrast, we observed WER reduction down to the block size of 240ms and 400ms on the CSJ \textit{dev} and \textit{test} sets, respectively.
A possible explanation is that the block-synchronous decoding introduced some effective inductive biases to the search process.

On the other hand, regarding LC-BLSTM and LC-Conformer encoders, we observed similar WERs on TEDLIUM2 and better WERs on CSJ, even with the frame-synchronous decoding.
This indicates that we can manage the display latency without WER degradation by leveraging future information in the encoder.

\begin{table}[t]
    \centering
    \begingroup
    \caption{Comparison of beam search type for MoChA with the ground-truth segmentation}\label{tab:result_beam_search}
    \vspace{-2mm}
    \scalebox{0.80}{
    \begin{tabular}{lcccccccc} \toprule
      \multirow{4}{*}{Encoder} & \multirow{4}{*}{\shortstack{Output\\synchronization}} & \multirow{4}{*}{$\tblock$} & \multicolumn{4}{c}{WER [$\%$] ($\downarrow$)} \\ \cmidrule(lr){4-7}
      & & & \multicolumn{2}{c}{TEDLIUM2} & \multicolumn{2}{c}{CSJ} \\ \cmidrule(lr){4-5} \cmidrule(lr){6-7}
      & &  & dev & test & dev & test \\ \midrule
       \multirow{7}{*}{UniLSTM} & Label & $\infty$ & 11.7 & 10.9 & 6.6 & 7.5 \\
    
       & Block & 64 & \bf{11.6} & 11.7 & \bf{6.4} & \bf{7.3} \\
       & Block & 40 & \bf{11.6} & 11.6 & \bf{6.4} & \bf{7.4} \\
       & Block & 32 & \bf{11.6} & 12.0 & \bf{6.5} & \bf{7.5} \\
       & Block & 24 & 11.8 & 12.2 & \bf{6.5} & 7.6 \\
       & Block & 16 & 12.0 & 12.6 & \bf{6.6} & 7.7 \\
       & Frame & 4 & 13.2 & 13.6 & 8.0 & 9.7 \\
       \midrule
       
      \multirow{3}{*}{LC-BLSTM} & Label & $\infty$ & 10.3 & \phantom{0}8.6 & 5.9 & 6.5 \\
      & Block & 40 & 10.4 & \phantom{0}\bf{8.6} & \bf{5.6} & \bf{6.3} \\
      & Frame & 4 & 10.4 & \phantom{0}\bf{8.6} & \bf{5.6} & \bf{6.3} \\
      \midrule

    \multirow{3}{*}{LC-Conformer} & Label & $\infty$ & \phantom{0}9.3 & \phantom{0}9.0 & -- & -- \\
    & Block & 32 & \phantom{0}9.4 & \phantom{0}\bf{8.9} & -- & -- \\
    & Frame & 8 & \phantom{0}\bf{9.3} & \phantom{0}\bf{9.0} & -- & -- \\
      \bottomrule
    \end{tabular}
    }
    \endgroup
    \vspace{2mm}
\end{table}

\begin{figure}[t]
  \centering
  \vspace{-4mm}
  \includegraphics[width=1.0\linewidth]{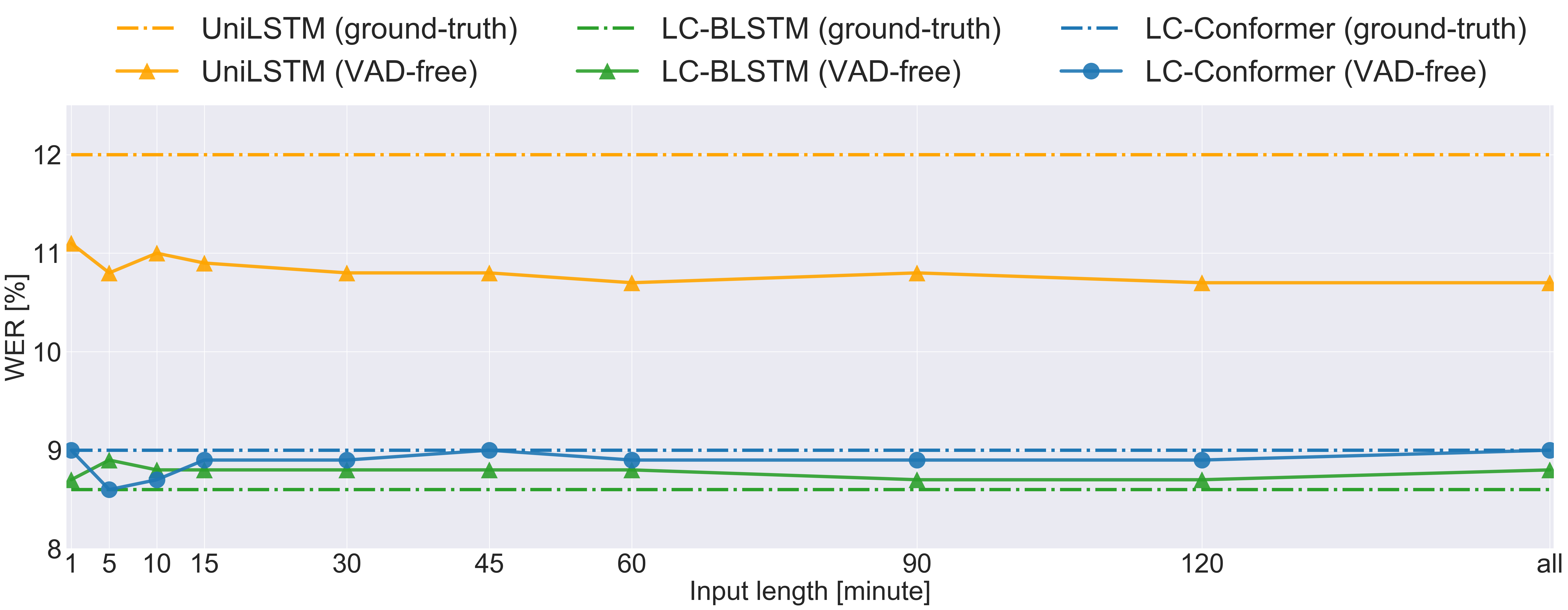}
  \vspace{-4mm}
  \caption{WER on simulated long-form recordings on the TEDLIUM2 test set. All models used block-synchronous decoding. The UniLSTM model used $\tblock = 32$.}
  \label{fig:longform_simulation_tedlium2}
  \vspace{2mm}
\end{figure}

\begin{table}[t]
    \centering
    \begingroup
    \caption{Ablation study with UniLSTM MoChA on simulated long-form recordings of TEDLIUM2 (all concatenated)}\label{tab:result_ablation}
    \vspace{-2mm}
    \scalebox{0.95}{
    \begin{tabular}{lccc} \toprule
    \multirow{2}{*}{Decoding} & \multicolumn{2}{c}{WER [$\%$] ($\downarrow$)} \\ \cmidrule(lr){2-3}
      & dev (1.59h) & test (2.61h) \\ 
      \midrule
      Block ($\tblock = 32$) & \bf{11.4} & \bf{10.7} \\
      \ \ w/o length normalization & 13.3 & 12.9 \\
      \ \ w/o LM state carryover & 11.6 & \bf{10.7} \\
      \ \ w/o safeguard & 19.7 & 19.7 \\
      \ \ w/o condition2 & 11.7 & 10.9 \\
      \ \ w/o back-off initialization & 12.9 & 12.0 \\
      \bottomrule
    \end{tabular}
    }
    \endgroup
\end{table}

\begin{table}[t]
    \centering
    \begingroup
    \caption{Session-level results with real long-form recordings}\label{tab:result_session}
    \vspace{-2mm}
    \scalebox{0.82}{
    \begin{tabular}{lcccccc} \toprule
      \multirow{4}{*}{Encoder} & \multirow{4}{*}{$\tblock$} & \multirow{4}{*}{VAD} & \multicolumn{3}{c}{WER [$\%$] ($\downarrow$)} \\ \cmidrule(lr){4-6}
      & & & \multirow{3}{*}{\shortstack{TED\\LIUM2\\test}} & \multirow{3}{*}{\shortstack{IWSLT\\tst20\{13/14/15\}}} & \multirow{3}{*}{\shortstack{CSJ\\test}} \\
      & &  &  &  &  \\
      & &  &  &  &  \\ 
      \midrule
    
      UniLSTM & 32 & \multirow{3}{*}{\xmark} & \bf{10.9} & {\bf 18.8}/{\bf 16.7}/{\bf 31.9} & \phantom{0}\bf{8.3} \\
      LC-BLSTM & 40 & & \phantom{0}\bf{8.8} & {\bf 16.9}/{\bf 15.1}/{\bf 29.7} & \phantom{0}\bf{8.7} \\
      LC-Conformer & 32 & & \phantom{0}\bf{8.9} & {\bf 16.4}/{\bf 15.1}/{\bf 30.1} & \phantom{0}-- \\
      \midrule
    
      UniLSTM & 32 & \multirow{3}{*}{\cmark} & 26.1 & 32.2/29.9/43.9 & 13.1 \\
      LC-BLSTM & 40 & & 18.3 & 25.5/23.0/37.1 & 11.9 \\
      LC-Conformer & 32 & & 34.9 & 37.3/36.2/43.5 & \phantom{0}-- \\

      \bottomrule
    \end{tabular}
    }
    \endgroup
    \vspace{2mm}
\end{table}

\vspace{-1mm}
\subsubsection{Long-form simulation with utterance concatenation}
\vspace{-1mm}
We next simulate variable lengths of long-form recordings to demonstrate the robustness of the proposed VAD-free inference algorithm.
We first sorted utterances in the evaluation set by the timestamps and concatenated adjacent utterances in a greedy way until the total length reached a certain length.
Therefore, multiple speakers could appear in a concatenated utterance.
We used $\tblock = 32$ for the UniLSTM encoder.
We present the results on TEDLIUM2 in Figure~\ref{fig:longform_simulation_tedlium2}.
The horizontal dot lines denote WERs of the block-synchronous decoding with the ground-truth segmentation.
We confirmed that the VAD-free inference was robust to long-form speech regardless of the encoder type and input lengths.
Interestingly, the performance of the UniLSTM encoder was significantly improved from the ground-truth segmentation for all the lengths.

\vspace{-1mm}
\subsubsection{Ablation study}
\vspace{-1mm}
We conduct the ablation study of the VAD-free inference algorithm on TEDLIUM2 in Table~\ref{tab:result_ablation}.
We concatenated all utterances in each evaluation set.
We observed that the safeguard had the largest impact, indicating that MoChA is more likely to generate $\langle$eos$\rangle$ at the end of blocks with the block-synchronous decoding.
Length normalization was also important for longer hypotheses to rank at the top.
Back-off initialization was important for the LSTM encoder to deal with frames around a reset point better.
Other techniques were slightly but consistently helpful.

\vspace{-1mm}
\subsubsection{Session-level evaluation on real long-form recordings}
\vspace{-1mm}
Finally, we conduct experiments on the real session-level lecture recordings.
Unlike the simulated experiments, there exist a lot of silence frames in the real recordings.
We compared the proposed method with the pre-segmentation with WebRTC VAD\footnote{\url{https://github.com/wiseman/py-webrtcvad}}.
Because both corpora do not necessarily contain all transcriptions in a session, we recognized all frames but removed tokens that did not match the ground-truth segments to calculate WER.
We also evaluated the models trained on TEDLIUM2 with the IWSLT \textit{tst2013}, \textit{tst2014}, and \textit{tst2015} sets~\cite{ansari2020findings}.
Results in Table~\ref{tab:result_session} showed that the proposed VAD-free inference achieved comparative WERs to those with the ground-truth segmentation on TEDLIUM2.
Although the results on CSJ degraded slightly compared to when using the ground-truth segmentation, they were much better than cascading VAD and ASR models.
Moreover, the degradation was much smaller than the CTC-based pre-segmentation~\cite{yoshimura2020end,fujita2020end}, which had a large latency to start recognition.
We observed that the VAD model was more likely to generate short segments that did not suit E2E models, especially for the LC-Conformer encoder.
This was because the LC-Conformer encoder had a total history context of 11.52 seconds.
Although there is room for improving VAD, the proposed unified framework eliminates the need for developing separate models independently.

\vspace{-1mm}
\section{Conclusions}
In this work, we have proposed the block-synchronous beam search decoding and the VAD-free inference algorithm to recognize unsegmented long-form speech with the hybrid CTC/MoChA framework.
Experimental evaluations on English and Japanese lecture corpora demonstrated that the proposed decoding method enabled stable recognition of long-form speech with a linear-time decoding complexity.
It was more accurate than performing VAD with an external model.

\footnotesize
\bibliographystyle{IEEEtran}
\bibliography{reference}

\end{document}